\documentclass[aps,prl,twocolumn,floatfix,superscriptaddress]{revtex4-2}
\usepackage{amsfonts,amsmath,amsthm,amssymb,amsopn,graphicx}
\usepackage{dsfont,bbm,xcolor}
\usepackage{enumitem}
\usepackage{hyperref}
\usepackage{breqn}
\usepackage{siunitx}
\usepackage{color}
\usepackage[]{breqn}
\usepackage[caption=false]{subfig}

\makeatletter
\let\cat@comma@active\@empty
\makeatother

\def\bra#1{\mathinner{\langle{#1}|}}
\def\ket#1{\mathinner{|{#1}\rangle}}

\newcommand{\ave}[1]{{\left\langle #1\right\rangle}}

\def\bra#1{\mathinner{\langle{#1}|}}
\def\ket#1{\mathinner{|{#1}\rangle}}

\def\red#1{\textcolor{red}{#1}}
\def\green#1{\textcolor{green}{#1}}
\def\blue#1{\textcolor{blue}{#1}}

\newcommand{\norm}[1]{\left\lVert#1\right\rVert}

\newcommand{\half}{{\textstyle\frac{1}{2}}}

\newcommand{\Eref}[1] {(\ref{#1})}

\usepackage{amsthm}
\theoremstyle{definition}

\begin{document}
\title{Dynamical l-bits in Stark many-body localization}
\author{Thivan M. Gunawardana}
\affiliation{Clarendon Laboratory, University of Oxford, Parks Road, Oxford OX1 3PU, United Kingdom}
\affiliation{Department of Mathematics, Imperial College London, London SW7 2AZ, United Kingdom}
\author{Berislav Bu\v{c}a}
\affiliation{Clarendon Laboratory, University of Oxford, Parks Road, Oxford OX1 3PU, United Kingdom}
\begin{abstract}
Stark many-body localized (SMBL) systems have been shown both numerically and experimentally to have Bloch many-body oscillations, quantum many-body scars, and fragmentation in the large field tilt limit. Likewise, they are believed to show localization similar to disordered MBL. We explain and analytically prove all these observations by rigorously showing the existence of novel algebraic structures that are \emph{exponentially} stable in time, which we call \emph{dynamical} l-bits.  Moreover, we show that many-body Bloch oscillations persist even at infinite temperature for exponentially long-times. We numerically confirm our results by studying the prototypical Stark MBL model of a tilted XXZ spin chain. Our work explains why thermalization was observed in a recent 2D tilted experiment. As dynamical l-bits are stable, localized and quantum coherent excitations, our work opens new possibilities for quantum information processing in Stark MBL systems.
\end{abstract}		

\maketitle
{\it Introduction ---} One of the seminal results of condensed matter physics was Anderson's discovery of localization of free electrons on a lattice \cite{Anderson}. Later it was shown that this localization can possibly persist even when the repulsive interactions between electrons cannot be neglected - a phenomenon dubbed many-body localization (MBL). 

One of the main results in MBL is its explanation in terms of exponentially localized intensive conservation laws called \emph{l-bits} \cite{lbits1,Serbyn}. The existence of these conservation laws blocks the flow of quantum information through the system. MBL has been numerically argued to lead to logarithmic entanglement growth \cite{ZnidaricMBL,MBLent2,MBLent3,MBLEnt4} and subdifussive transport \cite{PhysRevX.5.031032,PhysRevLett.114.160401,ZnidaricMBL2,PhysRevLett.114.100601,DaleyMBL,MBLexp} among other phenomena. Notably, MBL systems should be perfect insulators at any temperature. They have been the subject of huge study in recent decades (e.g. see the review \cite{Abanin_MBL})

However, the existence of disordered MBL has been somewhat controversial \cite{TomazMBL} despite certain exact \cite{Imbrie} and renormalization group \cite{JamirMBL1} results being offered. Rigorous results in many-body localization are therefore very important. 

Only very recently has another related form of MBL without disorder been demonstrated both theoretically and experimentally \cite{Stark1,Stark2,Stark3,guo2020stark}. This \emph{Stark} MBL occurs due to an external gradient field being added to an otherwise translationally invariant system. Related to well-known Bloch oscillations of non-interacting electrons,  Stark MBL (SMBL) demonstrates similar oscillations of various many-body observables in both numerics and experiments (e.g. \cite{Ribeiro,Stark2,Scherg,morong2021observation}). Likewise, recently both Hilbert space fragmentation \cite{Scherg,StarkFrag2} and quantum many-body scars \cite{PapicStark} have been numerically observed in these models. Fragmentation means that the Hamiltonian of the systems contains an exponential number of invariant subspaces \cite{Fragmentation1,Fragmentation2,ArnabFragmentation,fragmentationSanjay,SanjayNew} and quantum many-body scars are eigenstates that are equally spaced in energy and have low entanglement. Quantum many-body scars are known to imply oscillations from special initial states \cite{scars,scars9,scarsdynsym5,scarsdynsym6,scars5,scars6,SerbynScars}. 

In contrast to aforementioned Bloch oscillations in SMBL, disordered MBL systems do relax to stationary (time independent) states, albeit with a memory of their initial condition given by the l-bits \cite{PapicMBL,Abanin_MBL}. This stationarity of disordered MBL is in particular known to be present when expectation values of observables are averaged across disorder \cite{MBLdephasing1,MBL2}. This is puzzling because both forms of localization have an otherwise similar phenomenology \cite{Stark1}. 

In this Letter we rigorously prove that Stark MBL models have \emph{dynamical} l-bits that are \emph{exponentially} stable in time and are quasi-localized similarly to l-bits. The corresponding decay rate of the dynamical l-bits is given by the strength of the field gradient. These algebraic structures are distinct from both standard l-bits of disordered MBL and extensive dynamical symmetries \cite{Buca2}. We show that the existence of dynamical l-bits implies many-body Bloch oscillations even on the level of correlation functions at \emph{high} temperatures. The correlation functions persistently oscillate at frequencies given by the dynamical l-bits. We focus on the prototypical example of interacting electrons in an electric field gradient. We numerically confirm our theory by studying the infinite temperature correlation function and construct the dynamical l-bits. These dynamical l-bits explain the existence of quantum scars and fragmentation in  SMBL - both are shown to be consequences of the dynamical l-bits. Importantly, dynamical l-bits are quantum coherent and stable by construction and allow for storing qubits of information. Our work thus opens the possibility of quantum information storage and processing in SMBL systems. 


{\it Model ---}
We will focus on the following paradigmatic SMBL Hamiltonian,
\begin{multline}
    \label{model}
    H = J\sum_{r=1}^{L-1}(S_r^xS_{r+1}^x + S_r^yS_{r+1}^y) \\ + \Delta\sum_{r=1}^{L-1}{S}_r^z{S}_{r+1}^z + \sum_{r=1}^L \epsilon_r{S}_r^z
\end{multline}
where $L$ is the system size, $J$ is the hopping amplitude, $\Delta$ is the interaction strength (anisotropy) and $\epsilon_r$ is the external magnetic field at site $r$. This model of SMBL is equivalent to the fermionic model used in \cite{Ribeiro} via a Jordan-Wigner transformation, i.e. an interacting Wannier-Stark chain \cite{Wannier,AndreasStark1,AndreasStark2}. The non-interacting Wannier-Stark chain is well-known to feature Bloch oscillations \cite{BO1, BO2, BO3}. Similarly, in \cite{Stark1}, the field used was given by $\epsilon_r = Wr - \frac{\alpha r^2}{L^2}$ and it was shown that the inclusion of the quadratic component gave rise to a phase with MBL characteristics. It has also been shown that having a linear magnetic field $\epsilon_r = Wr$ also gives rise to an MBL-like phase beyond a critical tilt $W_c \approx 2.2$ in clean systems \cite{Nieuwenberg}. Supporting this, it has recently been shown that such oscillations exist even for moderate values of the tilt, but their origin is not well-understood \cite{Scherg}. We are interested in these oscillations and so we study a linear magnetic field for the rest of this Letter. However, we remark on the inclusion of a small quadratic potential in the Supplementary Material. As we will see, the exact form of the potential is not relevant for the qualitative conclusions. 


{\it Dynamical l-bits ---}
A fundamental theoretical advance in understanding disordered MBL is the discovery of l-bits \cite{Serbyn}. These are quasi-local operators which commute with the Hamiltonian and are thus conserved quantities. The existence of these implies that information about the initial state of the system must be preserved locally for all times and is thus accessible to local measurements. Hence the system cannot undergo thermal relaxation, leading to the MBL phase \cite{Abanin_MBL}.

We propose a similar theoretical framework for the SMBL phenomenon. As discussed previously, the key difference in SMBL is the observation of persistent many-body Bloch oscillations. To capture these, we look to the recently introduced concept of \textit{dynamical symmetries} \cite{Buca2}. These are defined to be \emph{extensive or local} spectrum generating algebras \cite{SGA} of ${H}$, i.e operators ${A}$ satisfying the relation \cite{footnote},
\begin{equation}
    \label{Dsym}
    [{H}, {A}] = \omega{A}
\end{equation}
where $\omega \neq 0$ is the \textit{frequency} of ${A}$. 

In this Letter, we extend the notion of dynamical symmetry to include \textit{dynamical l-bits} - these are operators satisfying \Eref{Dsym} which are similar to l-bits in MBL in the sense that they are quasi-localized (rather than strictly localized \cite{Buca}).

Now suppose that our system is initially (at $t = -\infty$) in thermal equilibrium and then locally perturbed suddenly at $t = 0$. This means that the perturbation takes the form $\delta(t){B}$ where {B} giving the new Hamiltonian ${H}' = {H} + \delta(t){B}$. According to standard results from linear response theory, the resulting deviation of the expectation value of an operator ${Q}$ at later times from it's equilibrium value is given by
    $\label{linresp}
    \ave{{Q}(t)}_{\rm{pert}} - \ave{{Q}} = -i\ave{[{Q}(t), {B}]}$
where $\ave{{Q}(t)}_{\rm{pert}} = \ave{e^{i{H}'t}{Q}e^{-i{H}'t}}$. If $\ave{AQ}\neq 0,\ave{AB}\neq 0$ for some dynamical symmetry $A$ \eqref{Dsym}, then $\ave{{Q}(t)}_{\rm{pert}}$ will oscillate forever with frequency $\omega$ \cite{Marko1}. More specifically, a Mazur lower bound on the amplitude of the oscillations exists \cite{Marko2}.

\begin{figure}
    \centering
    \includegraphics[width = \columnwidth]{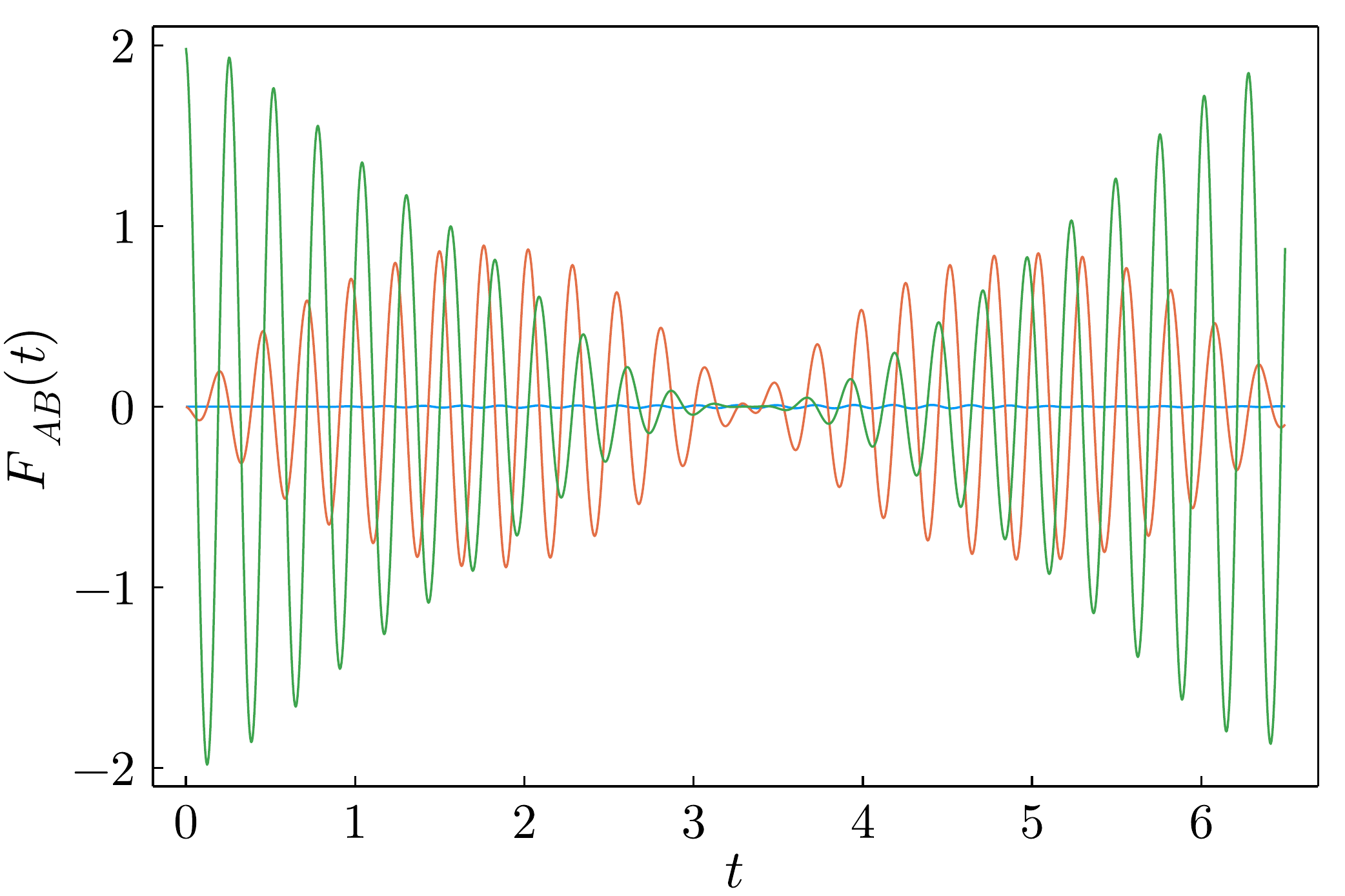}
    \caption{\textbf{Plot of the infinite temperature autocorrelation function in \Eref{autocorr} for various choices of local operators.} Here we consider three examples. The parameters used are $L = 16, W = 3, J = \Delta = 1$. In each case, we take the sudden perturbation at $t = 0$ to be ${B} = {S}^x_{L/2}$. The three choices of operators ${Q}$ are ${S}^+_{L/2}$ (\green{green}), ${S}^z_{L/2-1}{S}^+_{L/2}$ (\red{red}) and ${S}^z_{L/2+2}{S}^+_{L/2+3}$ (\blue{blue}).}
    \label{autocorrplot}
\end{figure}

We will focus on the infinite temperature case for which the relevant function from linear response theory to consider is the so-called fluctuation function given by,
\begin{equation}
    \label{autocorr}
    F_{QB}(t) = \half\ave{\left\{{Q}(t),{B}\right\}} = \ave{{Q}(t){B}}
\end{equation}
The above form is valid when ${Q}(0)$ and ${B}$ are traceless and this is precisely the case which interests us. 
We now numerically compute the autocorrelation function in \Eref{autocorr} for our model Hamiltonian \Eref{model} for a few pairs of operators  using DMRG \cite{itensor} (Fig.~\ref{autocorrplot}). Two cases show persistent many-body Bloch oscillations.  

 There is also one case where we observe no significant oscillations (the blue curve, which is almost flat). In this case, the time evolved operator acts on sites further away from the perturbation site. Thus, the memory of the perturbation is highly localized. These observations agree in general with how we expect dynamical l-bits would behave in autocorrelation functions, even though the oscillations are not of fixed amplitude. This leads us to theorize that the SMBL Hamiltonian possesses dynamical l-bits and that the operators in the plot which show oscillations have some finite overlap with them in the sense of \cite{Marko1}. What follows is the main result of this Letter where we prove that ${H}$ has exponentially stable dynamical l-bits, at least in the large tilt case.

{\it exponentially stable dynamical l-bits in SMBL ---}
We begin by noting that our Hamiltonian can be written in the form ${H} = {H}_{\rm{XX}} + {H}_{\rm{ZZ}} + {M}$, where 
    ${H}_{\rm{XX}} = J\sum_r({S}^x_r{S}^x_{r+1} + {S}^y_r{S}^y_{r+1}), \ 
     {H}_{\rm{ZZ}} = \Delta\sum_r({S}^z_r{S}^z_{r+1})$ and 
     ${M} = W\sum_r r{S}^z_r$.
Clearly, the eigenspectrum of ${M}$ is comprised of the values $\{W, 2W, ..., WL(L+1)/2\}$. Now define,
\begin{equation}
    \label{diagonal}
    {D} = \sum_n {P}_n({H}_{\rm{XX}} + {H}_{\rm{ZZ}}){P}_n
\end{equation}
where ${P}_n$ projects onto the eigenspace of ${M}$ with eigenvalue $Wn$.

We now assume that $W$ is large. It directly follows from the work of Abanin et al. \cite{pre-therm} (see also \cite{ElsePreth}) that there exists a quasi-local unitary operator ${Y}$ close to the identity such that up to exponentially long times $t^* \propto \exp{W}$,
\begin{equation}
    e^{i{H}t}{\mathcal{O}}e^{-i{H}t} \approx e^{i{Y}({D} + {M}){Y}^\dagger t}{\mathcal{O}}e^{-i{Y}({D}+ {M}){Y}^\dagger t}
\end{equation}
holds for any local operator ${\mathcal{O}}$. The error is exponentially small in $W$ (the reader is referred to \cite{pre-therm} for a more precise formulation of this statement along with a rigorous proof). In other words, up to exponentially long times, time evolution is governed by the effective Hamiltonian given by
${H}_{\rm{eff}} = {Y}{H}_{\rm{eff}}'{Y}^\dagger = {Y}({D} + {M}){Y}^\dagger$.

The aim now is to show that this effective Hamiltonian has dynamical l-bits by deriving exact dynamical l-bits for ${H}_{\rm{eff}}'$. To do this, we note that ${D}$ can be put in the more useful form \cite{pre-therm}
\begin{equation}
    \label{integral}
    {D} = \frac{1}{T}\int_0^T\text{d}t\ e^{i{M}t}({H}_{\text{XX}} + {H}_{\text{ZZ}})e^{-i{M}t}
\end{equation}
where $T = 2\pi/W$ (see \cite{SM}).
It is important to note that the reason the approaches \cite{pre-therm,ElsePreth} are applicable in SMBL, as opposed to disordered MBL is because the energy of the tilt is equally separated in SMBL, unlike the energy of the disordered field.

We then obtain that ${D} = {H}_{\rm{ZZ}}$ \cite{SM}, giving a rather simple form for ${H}_{\rm{eff}}'$ (cf. \cite{Fragmentation2,zisling2021transport}),
\begin{equation}
    \label{effective}
    {H}_{\text{eff}}' = {D} + {M} = \sum_r\left(\Delta{S}_r^z{S}_{r+1}^z + Wr{S}_r^z\right)
\end{equation}
This is simply an Ising Hamiltonian with a tilt! 

It is now easy to check that for $2\leq r\leq L-1$, the four operators given by
\begin{align}
    \label{exactdsym}
    {A}_1(r) &= {S}^+_r - 4{S}^z_{r-1}{S}^+_r{S}^z_{r+1} \nonumber \\
    {A}_2(r) &= {S}^z_{r-1}{S}^+_r - {S}^+_r{S}^z_{r+1} \nonumber \\
    {A}_3(r) &= {S}^+_r + 2{S}^z_{r-1}{S}^+_r + 2{S}^+_r{S}^z_{r+1} + 4{S}^z_{r-1}{S}^+_r{S}^z_{r+1} \nonumber \\
    {A}_4(r) &= {S}^+_r - 2{S}^z_{r-1}{S}^+_r - 2{S}^+_r{S}^z_{r+1} + 4{S}^z_{r-1}{S}^+_r{S}^z_{r+1}
\end{align}
 are exact strictly local dynamical l-bits of ${H}_{\text{eff}}'$ with corresponding frequencies
 \begin{equation}
     \omega_1 = \omega_2 = Wr,\ \omega_3 = Wr + \Delta,\ \omega_4 = Wr - \Delta
 \end{equation}

Now recall that the effective Hamiltonian for the SMBL model is actually ${H}_{\rm{eff}}$ rather than ${H}_{\rm{eff}}'$. However, for the above dynamical l-bits ${A}_j(r)$, we have that
    $[{H}_{\rm{eff}}, {Y}{A}_j(r){Y}^\dagger] = {Y}[{H}_{\rm{eff}}', {A}_j(r)]{Y}^\dagger = \omega_i{Y}{A}_j(r){Y}^\dagger$.
Since ${Y}$ is quasi-local, this gives a set of quasi-local dynamical l-bits for the effective Hamiltonian. Crucially, the corresponding frequencies remain unchanged. Furthermore, this effective Hamiltonian governs time evolution for the full Hamiltonian ${H}$ with exponentially small error for exponentially long times. Thus we conclude that the operators ${Y}{A}_j(r){Y}^\dagger$ are quasi-local dynamical l-bits of the full Hamiltonian which are valid as long as the effective Hamiltonian is valid.

Note that dynamical l-bits imply regular l-bits by the simple identity $[H,[A_j(r),A^\dagger_j(r)]]=0$, where $Q_j(r)=[A_j(r),A^\dagger_j(r)]$ is an exponentially localized l-bit. Unlike disordered MBL, dephasing here is not possible because, unlike the l-bits of disordered MBL, only four fundamental frequencies contribute to Stark dynamical l-bits, rather than a continuum in disordered MBL \cite{PapicMBL}. Importantly, this means that \eqref{effective} is the effective (up-to-exponentially long times) Hamiltonian in the l-bit basis that dictates dynamics, which has a much simpler form the disordered MBL one \cite{PapicMBL}. This also means that our l-bit basis is complete. 

{\it Numerical construction of the dynamical l-bits ---}
Even though we have shown the existence of dynamical l-bits for the full Hamiltonian, we have not found them explicitly since we do not know the operator ${Y}$. Our aim in this section is to numerically find these dynamical l-bits. To do this, we note that ${Y}$ is close to the identity, and so the exact dynamical l-bits of ${H}_{\rm{eff}}'$ are still highly relevant, and that the frequencies are unchanged. So we generalize the approaches of \cite{TomazNumerical} developed for conservation laws and we look at the operator given by,
\begin{equation}
    \label{tau}
    {\tau} = \lim_{T\rightarrow\infty}\frac{1}{T}\int_{-T}^{T}\text{d}t\ e^{-i\omega t}{U}^\dagger(t){\mathcal{O}}{U}(t)
\end{equation}
where $\omega \neq 0$ is a real number, $U(t) = e^{-iHt}$ is the time evolution operator and ${\mathcal{O}}$ is some strictly local operator. It can be shown that this satisfies the relation $[{H}, {\tau}] = \omega{\tau}$ regardless of ${\mathcal{O}}$ \cite{SM}. However, we require ${\tau}$ to also be (quasi) localized to be considered a dynamical l-bit and this is not necessarily satisfied for arbitrary local \emph{seed} operators ${\mathcal{O}}$.

\begin{figure}
\subfloat[]{
\includegraphics[width = 0.49\columnwidth]{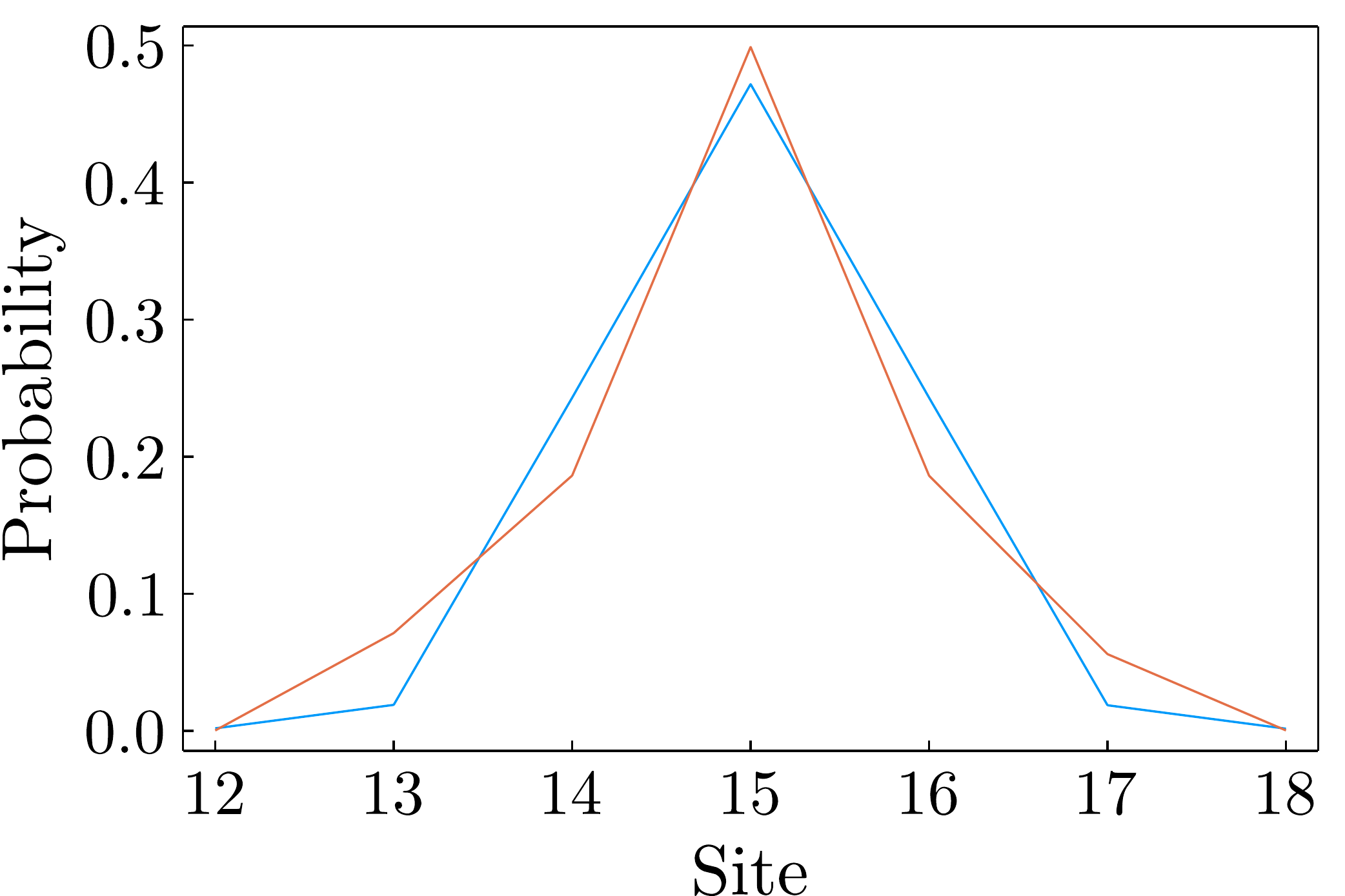}
}
\subfloat[]{
\includegraphics[width = 0.49\columnwidth]{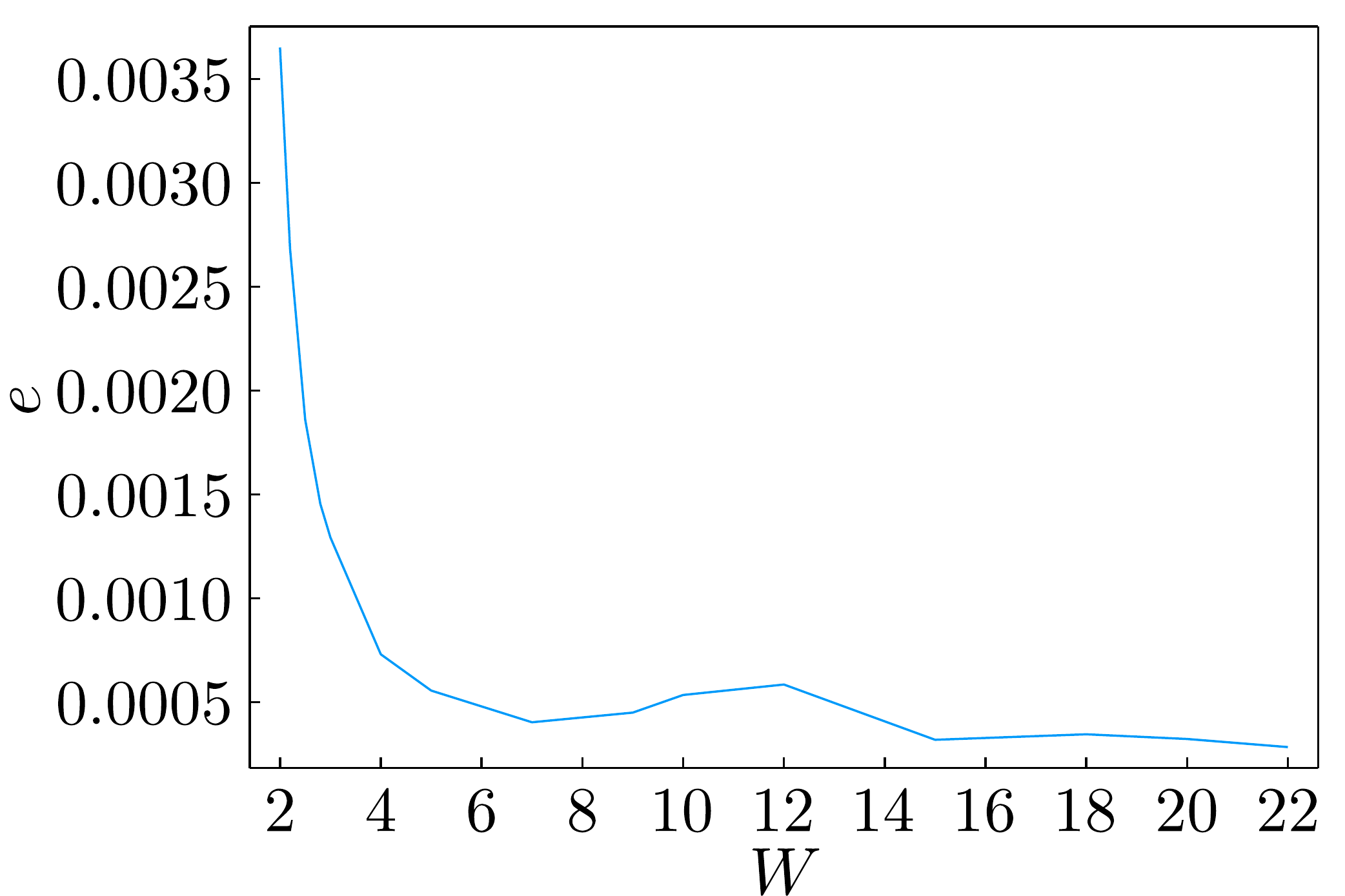}
}\\
\subfloat[]{
\includegraphics[width = 0.75\columnwidth]{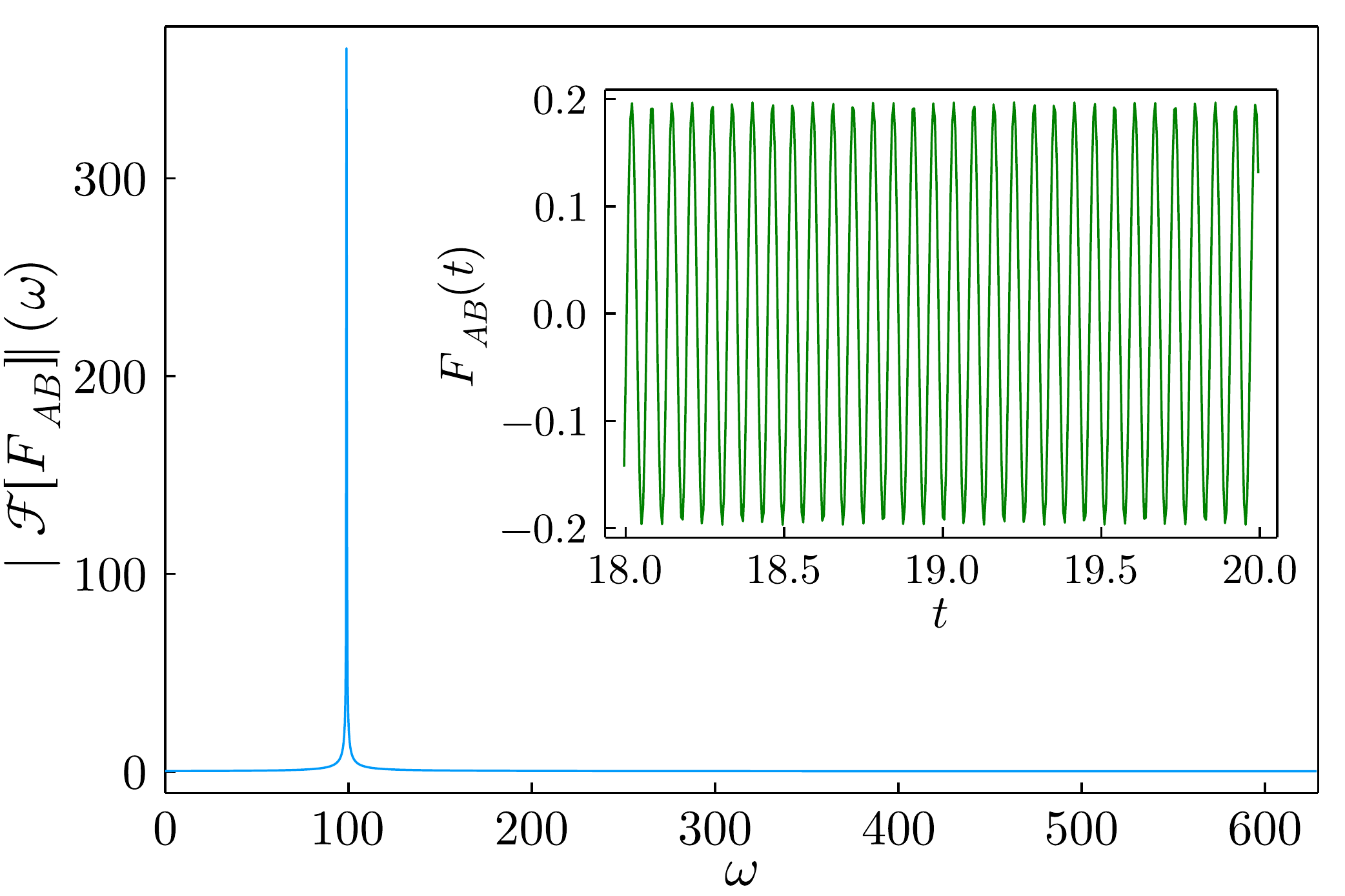}
}
\caption{\textbf{Results of the numerics on the dynamical l-bit candidate $\tau$.} The parameters used throughout are $J = \Delta = 1$ and $W = 10$ together with a total time $T = 350$ to simulate the limit $T \rightarrow \infty$ and timestep $dt = 0.005$ for the intergral in \Eref{tau}. The same timestep is also used to calculate the autocorrelation function. (a) Plots demonstrating locality of the operator $\tau$. The system size used is $L = 30$. The seed operators considered are $A_2(L/2)$ (\blue{blue}) and $A_4(L/2)$ (\red{red}). The probabilities give a measure of how much of ${\tau}$ lives on each site. As we can see, almost all of the operator is always concentrated on the central five sites, and the shape of the plot is not affected by the system size. (b) A plot showing the variation of the error $e$ with tilt strength $W$. We can see a high error for smaller values of $W$, where we naturally expect a lot of entanglement and thus larger errors due to truncation in DMRG. (c) Plot of the infinite temperature autocorrelation function (inset) $F_{AB}$ from \Eref{autocorr} with $A = S^x_{L/2}$ and $B = \tau$, and its Fourier transform for $L = 20$. The seed operator used was $A_4(L/2)$. The single spike in the Fourier transform confirms that $\tau$ is indeed a dynamical l-bit. }
\label{localityplot}
\end{figure}

By our previous arguments, we will use $\mathcal{O}={A}_j(r)$ as the seed operators, along with their corresponding frequencies. We again use DMRG \cite{itensor} to carry out these numerical simulations. 
We then determine locality by evaluating it's components in a \emph{complete} orthonormal product operator basis made up of products of (appropriately normalised) single site basis operators $\{{I}, {S}^+,{S}^-, {S}^z\}$. We then evaluate the total component for each site in the chain by sharing each component equally among the sites on which its corresponding basis operator lives (this is done simply to get the squares of the total site components to sum to $1$, and is not required to study locality). The squares of the total components for each site can be thought of as probabilities and these can be used to determine locality. The results as displayed in Fig.~\ref{localityplot} (a). Since the operator basis can be very large for large $L$, we find that it sufficient to only look at the basis operators acting on a radius of 3 about the site $L/2$.

These results make it clear that ${\tau}$ is quasi-local for these choices of seed operator, which agrees with our hypothesis. Thus we have numerically found quasi-local dynamical l-bits of the full Hamiltonian ${H}$. The accuracy of these can be looked at in a couple of ways. Finding the relative error in the commutator, given by
\begin{equation}
    e = \frac{\norm{[{H}, {\tau}] - \omega{\tau}}}{\norm{[{H}, {\tau}]}}
\end{equation}
shows that the error is only about 0.04\% which is very good. We can see how the error behaves when we change $W$ in Fig.~\ref{localityplot} (b) and we can see that it decays with increasing $W$, as expected. Furthermore, we can look at the autocorrelation function given by \Eref{autocorr}, with local operator ${A} = {\tau}$ and perturbation ${B} = {S}^x_{L/2}$ similar to what we did before. This is plotted in the inset of Fig.~\ref{localityplot} (c), where we can see clear, uniform oscillations at one frequency, which show that we have now found an accurate dynamical l-bit. The dynamical l-bit thus explains the existence of oscillations in Fig.~\ref{autocorrplot} at the appropriate frequency. This is confirmed by the Fourier transform in Fig.~\ref{localityplot} (c) and it demonstrates that the frequency of the oscillations is indeed the corresponding $\omega$ (which is 99 in this case). Moreover, we see that the operators which caused the non-uniform oscillations in Fig.~\ref{autocorrplot} do in fact overlap with the operators ${A}_i$ as conjectured previously. Furthermore, their support is indeed quasilocalized and close to the original dynamical l-bits as expected by the quasilocality of $Y$.  

{\it Quantum many-body scars and fragmentation ---} Quantum many-body scars \cite{scars2,scars,scars9,scarsdynsym5,scarsdynsym6,scars5,scars6,SerbynScars,scarsPnew}, which are an (at least) extensive number of eigenstates that have low entanglement. The are related to oscillations from special initial states. These states have recently been numerically identified in Stark MBL models \cite{PapicStark}. Our analytically proven dynamical l-bits directly imply quantum scars \cite{SanjayReview,scarsdynsym4,scarsdynsym5,scarsdynsym6} (see also Suppl of \cite{Buca2} for an earlier proof implying scarring and that includes dissipation). In our case, the existence of scars follows from the fact that dynamical l-bit when acting on the (product) ground state $\ket{0}$ will create eigenstates with low-entanglement that are equally separated in energy, e.g.
\begin{equation}
H\left(\prod_{\{r_i\}}A_j(r_i) \ket{0}\right)= d(j) \omega_j \left(\prod_{\{r_i\}}A_j(r_i) \ket{0}\right)
\end{equation}
where $[H,A_j(r_i)]=\omega_j A_j(r_i)$, and $d(j)$ is the number of $A_j(r_i)$ that appear in the product.  We set $H\ket{0}=0$ for simplicity. The entanglement of the state is guaranteed to be low due to the localized structure of $A_j(r_i)$.

We note that dynamical l-bits imply quantum scars, but not the other way around. More specifically, models with quantum many-body scars have oscillations only for very special initial states, whereas dynamical l-bits imply oscillations generically and even at infinite temperature as shown here. 

Fragmentation follows immediately from the dynamical l-bits by the arguments presented in \cite{strictlylocalfrag}. This is likewise consistent with the results of \cite{StarkFrag2}. In fact this means that Stark MBL fragmentation is not true fragmentation \cite{Fragmentation1,Fragmentation2,statisticallocalization}, but rather \emph{local fragmentation} as defined by \cite{strictlylocalfrag}. 

{\it Conclusion ---}
In this Letter we studied SMBL and investigated the origin of persistent oscillations which have been numerically and experimentally observed \cite{Stark1, Nieuwenberg, Ribeiro,morong2021observation}. We argued that this is due to the existence of dynamical l-bits, similar to how the MBL phenomenon has been theoretically described by normal l-bits \cite{Serbyn}. We then proved that in the large tilt case, the SMBL Hamiltonian can be reduced to an effective Hamiltonian up to exponentially long using a theorem from \cite{Abanin_MBL} and further showed the existence of exact and complete dynamical l-bit basis of this effective Hamiltonian. Thereafter, we used these to numerically construct dynamical l-bits of the full Hamiltonian with excellent accuracy. A similar effective Hamiltonian was obtained in a very recent preprint (up to a rotating wave basis transform) in \cite{zisling2021transport} where transport was studied (cf. also \cite{Fragmentation2}). However, here we focused on oscillations, scars and fragmentation, as well as found the complete l-bit basis. We have proven that dynamical l-bits imply persistent oscillations in the autocorrelation function, even at infinite temperature, quantum many-body scars, and Hilbert space fragmentation, as well as l-bits. The fact that the dynamical l-bits of SMBL have only four fundamental frequencies explains why SMBL has many-body Bloch oscillations, unlike disordered MBL \cite{PapicMBL}. Even though the XXZ model we studied is distinct from the tilted Fermi-Hubbard models with scars and fragmentation, our results indicate that these models likewise have dynamical l-bits. Note that dynamical l-bits immediately imply many-body flat bands \cite{Kuno} in the tilted XXZ spin chain \cite{Buca}. Our approach relies on the tilt being single body and therefore a tilt with with two-body terms is expected to thermalize, which is fully consistent with the 2D experiment of \cite{Stark2D}. Moreover, we predict that the putting a two-body tilt in both directions will not cause absence of thermalization in contrast to the proposal in \cite{Stark2D}.

Our work opens many new avenues for future work. In future work we will study in Stark MBL models possible realizations of time crystals in both driven (discrete) \cite{ElseTC,Chinzei,Chinzei2,UedaTC,timecrystal2,timecrystal3,LazaridesDissipation,guo2020detecting,stochasticdiscrete,dissipativeTCobs} (cf. also \cite{StarkTC}) and dissipative models \cite{Buca2,Esslinger,Esslinger2,Shammah,Booker_2020,Orazio,Jamir1,Cosme1,Cosme2,Jamir2,Fazio,liang2020time,Lesanovsky,Seibold,Hadiseh}, synchronization \cite{buca2021algebraic,SynchronizationNEW,QSynch}, and other possible kinds of non-stationary dynamics \cite{Kollath,versteeg2020nonequilibrium,Benitez1,Benitez2,versteeg2020nonequilibrium,Piazza,Sarang,mendozaarenas2021selfinduced,BucaJaksch2019,metastabilityNEW,Dora,guarnieri2021time,Pozsgay,Carlos,etapairing}. Likewise, connection with large-tilt and large interactions limits in 1D models will be explored \cite{LenartFolded1,LenardFolded2,tartaglia2021realtime}. Most intriguing, however, is the fact that a dynamical l-bit is local coherent excitation that can store a qubit. This hints that Stark MBL systems could have potential for robust quantum information storage and processing. 

\begin{acknowledgments}
{\it Acknowledgements ---} We thank Z. Papi\'{c} and F. Pollmann for useful discussions, A. Lazarides for pointing out a recent preprint \cite{zisling2021transport} and P. Sala for numerous enlightening suggestions and useful discussion. Funding from EPSRC programme grant EP/P009565/1, EPSRC National Quantum Technology Hub in Networked Quantum Information Technology (EP/M013243/1) is gratefully acknowledged. T.M.G. acknowledges funding from a Departmental Roth Scholarship from the Department of Mathematics, Imperial College London for part of this work.
\end{acknowledgments}

\bibliographystyle{apsrev4-2}
\bibliography{main}

\widetext
\clearpage
\begin{center}
\textbf{\large Supplementary Material: Dynamical l-bits in Stark many-body localization}\\
\end{center}
\setcounter{equation}{0}
\setcounter{figure}{0}
\setcounter{table}{0}
\setcounter{page}{1}
\makeatletter
\renewcommand{\theequation}{S\arabic{equation}}
\renewcommand{\thefigure}{S\arabic{figure}}

In this Supplementary Material, we give more details on some of the assertions made in the main text. 

\section{A--- Derivation of the integral form of $D$}
In this section, we show the equivalence of \Eref{diagonal} and \Eref{integral} in the main text. Let $\ket{m}$ and $\ket{n}$ be eigenstates of $M$ with corresponding eigenvalues $E_m$ and $E_n$ respectively. Then, starting from \Eref{integral}, we have the matrix element
\begin{align}
    \label{melements}
    \bra{n}D\ket{m} &= \frac{1}{T}\int_0^T \textrm{d}t\ \bra{n}e^{iMt}(H_{\text{XX}} + H_{\text{ZZ}})e^{-iMt}\ket{m} \nonumber \\
    &= \frac{1}{T}\int_0^T \textrm{d}t\ e^{i(E_n - E_m)t}\bra{n}(H_{\text{XX}} + H_{\text{ZZ}})\ket{m} \nonumber \\
    &= \delta_{E_n, E_m}\bra{n}(H_{\text{XX}} + H_{\text{ZZ}})\ket{m}
\end{align}
where the final equality follows from the fact that $T = 2\pi/W$ and that the eigenvalues of the tilt $M$ are all integer multiples of $W$. The matrix elements given by \Eref{melements} are non-zero only if $E_n = E_m$, that is, we are within a single eigenspace of $M$. Furthermore, when we are inside a single eigenspace of $M$, \Eref{melements} gives us the correct matrix elements of \Eref{diagonal}. Since the eigenstates we picked were arbitrary, this proves the equivalence of the two expressions for $D$.

\section{B--- Full Derivation of the transformed effective Hamiltonian $H_{\text{eff}}'$}
In this section, we show that $H_{\text{eff}}'$ is indeed given by the tilted Ising model in \Eref{effective} in the main text. We define the operator
\begin{equation}
    K = \frac{J}{2}\sum_r {S}^+_r{S}^-_{r+1}
\end{equation}
where ${S}^{\pm}_r$ are the spin raising and lowering operators respectively. It can then be easily observed that $H_{\text{XX}} = K + K^\dagger$. Furthermore, we can use the commutators $[S^z_r, S^{\pm}_{r'}] = \pm S^{\pm}_r\delta_{rr'}$ to obtain the closed algebra given by (cf. \cite{Bloch1,AndreasStark1,AndreasStark2})
\begin{equation}
    [M, K^\dagger] = WK^\dagger,\ [M, K] = -WK,\ [M, H_{\text{ZZ}}] = 0.
\end{equation}
This shows that $K$ and $K^\dagger$ are dynamical symmetry operators of $M$, which is not very interesting by itself, but does allow us to utilise \Eref{Dsym} from the main text to evaluate the integral in \Eref{integral}. Indeed, we have
\begin{align}
    D &= \frac{1}{T} \int_0^T \textrm{d}t\ e^{iMt}(K + K^\dagger + H_{\text{ZZ}}) e^{-iMt} \nonumber \\
    &= \frac{1}{T} \int_0^T \textrm{d}t \left(e^{iWt}K^\dagger + e^{-iWt}K + H_{\text{ZZ}}\right)\nonumber \\
    &= \frac{1}{2\pi i}\left[e^{iWt}K^\dagger + e^{-iWt}K + H_{\text{ZZ}}\right]_0^{2\pi/W} + H_{\text{ZZ}} \nonumber \\
    &= H_{\text{ZZ}}
\end{align}
Then, we get that
\begin{equation}
    H_{\text{eff}}' = D + M = \sum_r (\Delta{S}^z_r{S}^z_{r+1} + Wr{S}^z_r)
\end{equation}
as claimed.
\newpage
\section{C--- Proof that $\tau$ is a dynamical symmetry operator}
In this section we show that the operator $\tau$ given by \Eref{tau} does indeed satisfy the commutation relation $[H, \tau] = \omega\tau$. Let the eigenbasis of $H$ be given by $\{\ket{n}\}$ with corresponding eigenvalues $E_n$. Then, we have
\begin{align}
    [{H},{\tau}] &= \lim_{T\rightarrow\infty}\frac{1}{T}\int_{-T}^{T}\text{d}t\ e^{-i\omega t}{U}^\dagger(t)[{H},{\mathcal{O}}]{U}(t) \nonumber \\
    &= \lim_{T\rightarrow\infty}\frac{1}{T}\sum_{m,n}\int_{-T}^{T}\text{d}t\ e^{i((E_m - E_n) - \omega)t}(E_m - E_n)\ket{m}\bra{m}{\mathcal{O}}\ket{n}\bra{n} \nonumber \\
    &= \omega\left(\lim_{T\rightarrow\infty}\frac{1}{T}\sum_{m,n}\int_{-T}^{T}\text{d}t\ e^{i((E_m - E_n) - \omega)t}\ket{m}\bra{m}{\mathcal{O}}\ket{n}\bra{n}\right) \nonumber \\
    &= \omega{\tau}
\end{align}
where the third equality follows from the fact that the integral over time of the complex exponential is zero unless $E_m - E_n = \omega$. The final equality can be seen by working backwards on $\tau$ by introducing resolutions of the identity similarly to what we did for $[H, \tau]$ above. This shows that $\tau$ is in fact a dynamical symmetry operator of $H$ for arbitrary local operators $\mathcal{O}$.

\section{D--- Including a small quadratic component to the magnetic field}
Here we consider what happens if we add a small quadratic potential to the external magnetic field. This means that we now consider $\epsilon_r = Wr + \frac{\alpha j^2}{L^2}$ where $\alpha \ll W$. In this case, $H = H_{\text{XX}} + H_{\text{ZZ}} + M$ just like before, with $H_{\text{ZZ}}$ now changed to
\begin{equation}
    H_{\text{ZZ}} = \sum_r\left(\Delta{S}^z_r{S}^z_{r+1} + \frac{\alpha r^2}{L^2}{S}^z_r\right)
\end{equation}
Since this also commutes with $M$, the arguments in section B still hold and we now get that
\begin{equation}
    H_{\text{eff}}' = D + M = \sum_r \left(\Delta{S}^z_r{S}^z_{r+1} + \left(Wr + \frac{\alpha r^2}{L^2}\right){S}^z_r\right)
\end{equation}
Note that we can still use the results by Abanin et al. because $\alpha \ll W$ and so $W$ is indeed large relative to the local energy scales of $H_{\text{XX}} + H_{\text{ZZ}}$. Thus it is easy to see that we end up with the same dynamical l-bits as before, but with the frequencies incremented by the quadratic potential at the central site of the dynamical l-bit.
\end{document}